\documentclass{llncs}
\usepackage{amssymb}
\usepackage{amsmath,amssymb}
\usepackage{enumerate}
\usepackage{graphics,graphicx}
\usepackage{color}

\usepackage{xspace}
\usepackage{bbm}
\usepackage{paralist}
\usepackage{longtable}

 \usepackage[english]{babel}
\usepackage[ruled,vlined,linesnumbered]{algorithm2e}
\usepackage[utf8]{inputenc}
\RequirePackage{pstricks,pst-tree,pst-node} 
\RequirePackage{pst-plot}
\newcommand*{\itemb}{\item[$\bullet$]}
\newcommand*{\fonction}[5]{ #1 \text{ }\mathrm{ : }\text{ } \left\{ \begin{split}#2 \quad &\to \quad #3 \\ #4 \quad &\mapsto \quad #5 \\ \end{split}\right. }
\newcommand*{\dsum}[2]{\displaystyle{\sum_{#1}^{#2}} }
\newcommand*{\dcup}[2]{\displaystyle{\underset{#1}{\overset{#2}{\bigcup}}}}
\newcommand*{\llbracket}{[\![}
\newcommand*{\rrbracket}{]\!]}

\renewcommand*{\bar}[1]{\overline{#1}}

\usepackage{color}
\RequirePackage{mathrsfs}
\DeclareMathOperator{\opt}{opt}
\DeclareMathOperator{\argmin}{argmin}
\DeclareMathOperator{\argmax}{argmax}
\DeclareMathOperator{\select}{select}
\DeclareMathOperator{\order}{order}
\DeclareMathOperator{\assigned}{assigned}
\DeclareMathOperator{\true}{true}

\DeclareMathOperator{\anyvertex}{anyvertex}
\DeclareMathOperator{\size}{size}
\DeclareMathOperator{\neighbours}{neighbours}
\DeclareMathOperator{\clusters}{clusters}

\title{Next Generation Cluster Editing}
\author{Thomas Bellitto\inst{1}\thanks{This work was done during an
    internship of the first author at CWI.}  \and Tobias Marschall\inst{2} \and\\
  Alexander Sch\"onhuth\inst{2} \and Gunnar W.\ Klau\inst{2}}

\institute{
Department of Computer Science and Telecommunications,\\ENS
Cachan/Rennes, France \and 
 Life Sciences, Centrum Wiskunde \& Informatica (CWI),\\ Amsterdam, The
Netherlands}

\begin{document}








 \maketitle

\begin{abstract}
This work aims at improving the quality of structural variant
prediction from the mapped reads of a sequenced genome. We suggest a new model based on cluster editing in
weighted graphs and introduce a new heuristic algorithm that allows to
solve this problem quickly and with a good approximation on the huge
graphs that arise from biological datasets.

\medskip

\end{abstract}


\section*{Introduction}

Structural variations in the human genome play a key role in genetic diversity. Many diseases have been associated with genetic alterations and locating them can be valuable for developing appropriate treatments. Thus, analysing and understanding human genetic variations has become a key issue in many fields of medicine and biology, and the advent of next-generation sequencing enabled the creation of huge databases on human genomes. However methods for exploiting the data and detecting the structural variations have not kept up with sequencing. There is ample evidence that existing datasets contain variations that the current methods cannot discover. 

Many variant finders were developed in recent years to address this problem, but discovering small insertions and deletions is still a very challenging problem. Our works are based on Clever (CLique Enumerating Variant findER), a tool introduced in 2012 in \cite{Clever} that achieves very good results and compares favourably to the other state-of-the-art tools. However, Clever is based on clique enumerating, a model that leads to problems we know how to solve exactly but that shows some flaws. In order to improve the results of Clever, we designed a new approach based on a different optimisation problem: cluster editing. The new model seems more appropriate but leads to a difficult problem that has to be solved on huge data structures.

In the first part of the report, we will describe the principle and the work scheme of Clever and point out the problems we want to solve and that justify the cluster editing model. We will then describe the state of the art on cluster editing and expand on a particular class of graphs: point graphs.

The second part presents our contribution to the new model. We study exact and heuristic algorithms to solve cluster editing in weighted point graphs. We then use it to design a fast heuristic algorithm suited to biological instances. We finally present experimental results to validate the model and the heuristic.

We believe that the validation of this new model and the theoretical study that has been undertaken can give a new direction for future research regarding Clever and can lead to significant improvements of the results.

\section{Context}

\subsection{Clever}

\subsubsection{Principle of Clever}

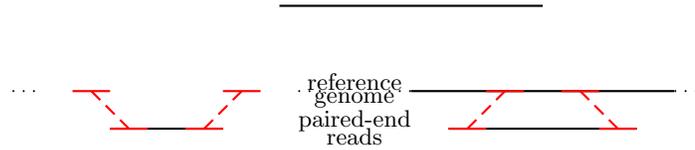
\begin{figure}[!h]
\psline{-}(0,0.5)(3.5,0.5)
\centering\[\begin{pspicture}(6.5,0.8)
\psline[linestyle=dotted](-0.3,0.5)(0,0.5)
\psline[linestyle=dotted](3.5,0.5)(3.8,0.5)
\psline[linecolor=red](0.5,0.5)(1,0.5)
\psline[linecolor=red](2.5,0.5)(3,0.5)
\psline{-}(1,0)(2.5,0)
\psline[linecolor=red](1,0)(1.5,0)
\psline[linecolor=red](2,0)(2.5,0)
\psline[linecolor=red, linestyle=dashed](0.75,0.5)(1.25,0)
\psline[linecolor=red, linestyle=dashed](2.75,0.5)(2.25,0)

\psline{-}(5,0.5)(8.5,0.5)
\psline[linestyle=dotted](4.7,0.5)(5,0.5)
\psline[linestyle=dotted](8.5,0.5)(8.8,0.5)
\psline[linecolor=red](6,0.5)(6.5,0.5)
\psline[linecolor=red](7,0.5)(7.5,0.5)
\psline{-}(5.5,0)(8,0)
\psline[linecolor=red](5.5,0)(6,0)
\psline[linecolor=red](7.5,0)(8,0)
\psline[linecolor=red, linestyle=dashed](5.75,0)(6.25,0.5)
\psline[linecolor=red, linestyle=dashed](7.75,0)(7.25,0.5)

\rput(4.25,0.1){\footnotesize{\text{paired-end}}}
\rput(4.25,-0.1){\footnotesize{\text{reads}}}
\rput(4.25,0.62){\footnotesize{\text{reference}}}
\rput(4.25,0.38){\footnotesize{\text{genome}}}

\end{pspicture}\]
\caption{The sequenced fragments we align are displayed in red. According to their sizes, alignments may indicate deletions (left) or insertions (right).}
\label{read}
\end{figure}

A paired-end read is a fragment of DNA whose extremities have been sequenced. We then look for those sequences in a reference genome to align the paired-end read on the genome (see figure \ref{read}). 

According to their sizes, alignments are more or less likely to indicate insertions or deletions, and the positions of the reads on the genome will help us locating the variations. 

Given a dataset that usually consists of millions of reads, Clever finds variations by processing as described:

\begin{enumerate}
 \item Clever organises those reads in what we call a read alignment graph. Clever then draws edges between reads who are likely to stem from the same allele and enumerates maximal cliques, that represent contradiction-free groups of reads. More details about read alignment graphs and the clique enumerating algorithm are given in the next part.
 \item  At this stage, we have many cliques that we assume to contain reads that stem from the same allele. We compute for each clique a $p$-value that indicates how likely it is to observe those reads under the hypothesis that there were no insertions and a $p$-value with the hypothesis that there were no deletions. Predictions will therefore come from the lowest $p$-values.
 \item Finally, Clever uses the Benjamini-Hochberg process \cite{fdr} to control false discovery rate: if we have $c$ clusters, to control the false discovery rate at $r$, we will return the biggest number $p$ of predictions such that all their $p$-values are under $\displaystyle{\frac{rp} c}$. As we use different formulas to compute the $p$-values of insertions and deletions, Clever will apply this process for the insertions and the deletions separately. In practice, we often choose a 10\% false discovery rate.
\end{enumerate}

\subsubsection{Read Alignment graph}

The read alignment graph is a graph where each vertex represents a read and where we draw edges between statistically "close" reads. To draw an edge, we want the two following conditions to hold:
\begin{itemize}
 \itemb the reads must be close in position (once they are aligned on the reference genome), meaning that the overlap between the reads must be big enough
 \itemb the sizes of the reads must be close enough.
\end{itemize}

\vspace{0.35cm}
Formally, we call $I(A)$ the length of a read $A$. In practice, we can assume that the length of the reads are normally distributed. We call $\mu$ the mean value and $\sigma$ the standard deviation.

Let $A$ and $B$ be two reads:
\begin{itemize}
 \itemb We call $\Delta(A,B)=|I(A)-I(B)|$ the length difference
 \itemb We call $O(A,B)$ the size of the overlap
 \itemb We call $\bar I(A,B)=\displaystyle{\frac {I(A)+I(B)} 2}$ the mean lengths
 \itemb We define $U(A,B)=\bar I(A,B)-O(A,B)$
\end{itemize}

\vspace{0.35cm}
Let $X\sim \mathcal N_{(0,1)}$. If $A$ and $B$ are two different overlapping reads, we will draw an edge between $A$ and $B$ iff:
\[\left\{\begin{split}
 \mathbb P \left(|X|\geqslant	\frac 1 {\sqrt 2} \frac{\Delta(A,B)}{\sigma}\right)&\leqslant T\quad\text{(size criteria)}\\
 \mathbb P \left(X\geqslant \sqrt 2 \frac{U(A,B)-\mu}{\sigma}\right)&\leqslant T\quad\text{(overlap criteria)}\\
\end{split}\right.\]

where $T$ is a threshold value that will determine the graph density.

We can also use those criteria to weight our edges. The weight $w(A,B)$ of the edge between $A$ and $B$ is given by the formula:

\[\begin{split}
 &w_{\text{size}}(A,B)=\ln T - \ln\left(\mathbb P \left(|X|\geqslant \frac 1 {\sqrt 2} \frac{\Delta(A,B)}{\sigma}\right)\right)\\
 &w_{\text{overlap}}(A,B)= \left\{\begin{split} &-\infty\quad\text{ if }O(A,B)=0\\ &\ln T - \ln\left(\mathbb P \left(X\geqslant \sqrt 2 \frac{U(A,B)-\mu}{\sigma}\right)\right)\text{ if }O(A,B)\geqslant 0\end{split}\right. \\
 &w(A,B)=\min(w_{\text{size}}(A,B),w_{\text{overlap}}(A,B))
\end{split}\]

As most of the reads do not overlap at all (actually, the proportion of overlapping reads is $O(\frac 1 n$)), the presence/absence of most of the edges (unweighted case) or their weight will often be determined by the overlap. This makes unweighted read alignment graphs look like interval graphs, a well-studied class of graphs.

\vspace{0.35cm}
\begin{definition}[interval graph]
Given a set $\mathcal I$ of intervals, the associated interval graph is the graph $G=(\mathcal I, \{(A,B)\in \binom {\mathcal I} 2 \mid A\cap B \neq \varnothing\})$. We represent each interval with a vertex and we draw edges between overlapping intervals.
 \end{definition}

There are $O(n\*k)$-time exact algorithms for the maximum clique enumerating problem in interval graphs, where $k$ is a bound on the size of the neighbourhood of the vertices and can in practice be assumed constant. As reads that overlap do not have to be connected in read alignment graphs because of size difference, read alignment graphs are only interval graphs with missing edges, which is a trivial property as cliques are interval graphs. However, even if it is difficult to write it formally, at low coverage, read alignment graphs will be "close" to interval graph and exact algorithms that run in polynomial time for interval graphs will still behave polynomially. 

Here is the algorithm used by Clever: (we use the notation $N(A)$ to denote the neighbourhood of $A$)

\vspace{0.35cm}
\begin{algorithm}[H]
let $L$ be the sorted list of the $2n$ endpoints of the reads

$\mathscr C\leftarrow \varnothing$ will be the set of active cliques

\For{$p\in L$}{
\eIf{$p$ is the left endpoint of a read $A$}{
\eIf{$\forall C\in\mathscr C, N(A)\cap C=\varnothing$}
{$\mathscr C\leftarrow\mathscr C\cup \{A\}$}
{\For{$C\in\mathscr C$}{
\If{$C\cap N(A)=C$}
{$C\leftarrow C\cup\{A\}$}
\If{$C\cap N(A)=\varnothing$}
{$\mathscr C\leftarrow \mathscr C\cup\{(C\cap N(A))\cup\{A\}\}$}
}
}
}
{$p$ \text{is the right endpoint of a read} $A$:
{\For{$C\in\mathscr C$}{
\If{$A\in C$}
{$\mathscr C\leftarrow \mathscr C\setminus C$

return $\mathscr C$}
}
}
}
}
\caption{Clique enumerating \cite{Clever}}
\end{algorithm}

It is established in \cite{Clever} that this algorithm runs in time $O(n(\log n+kc^2)+s)$ where $k$ is an upper bound on the size of the cliques, $c$ is an upper bound on the size of the set of active cliques, and $s$ is the size of the output. As we said previously, every graph could be seen as an interval graph with missing edges, but in the general case, $c$ cannot be bounded with a polynomial expression. However, with interval graphs, $c\leqslant k$ and in practice, we can assume that with our read alignment graphs, $c=O(k)$.

\subsubsection{Limits of Clever}

First of all, let us take a closer look at the model. We say that two sets of vertices overlap iff their intersection is non empty. Clever makes predictions on groups of reads that are supposed to stem from the same allele, thus, it makes no sense for two different groups to overlap, but we form those groups by enumerating maximal cliques that will always overlap. 

Just like reads, predictions are said to overlap if and only if both starts before the other end. Thus, it does not make sense to predict, for example, two overlapping insertions in a genome. However, the predictions of Clever may contain overlapping insertions or deletions.

We still can use a post-processing script that deletes predictions that overlap with others until all the predictions are disjoint. However, dropping information will damage the quality of the predictions: what Clever optimises is the file of overlapping predictions and not the post-processed file. Ultimately, we would like to find a model that avoids overlapping predictions or at least, minimises them so that the optimised file (with overlap) is as close as possible to the final post-processed file.

Another completely different problem arises when we try to run Clever on higher coverage data, e.g. to analyse viral populations (see \cite{Virus}). The $O(n(\log n+kc^2)+s)$ complexity will no longer be acceptable as $k$ and $c$ will reach values that are too high.

\subsection{Optimal cluster editing problem}

\subsubsection{Definition}

\paragraph{Unweighted case} ~

Given an undirected unweighted graph $G=(V,E)$, the cluster editing problem consists of transforming $G$ into a clustered graph $G'$, \textit{i.e.} into a union of disjoint cliques by adding or removing edges. 

A solution to this problem can be represented by the set $R\subset E$ of edges to remove and the set $A\subset \binom V 2 \setminus E$ of edges to add, or alternatively by a partition of $V$ showing which clusters have been made.

The cost of a solution is the number of modifications made on $G$: $|A|+|R|$.

The optimal cluster editing problem consists in finding the solution with the smallest cost.

\vspace{0.35cm}
\begin{example}
 Let $G$ be the graph shown in figure \ref{excep1}.
\begin{figure}[!h]
\begin{minipage}{0.32\linewidth}

\[\begin{psmatrix}[nodesep=5pt,rowsep=0.8cm,colsep=1cm] 
1&2&\\
3&4&5\\
&6&7
\ncline{1,1}{1,2}
\ncline{1,2}{2,1}
\ncline{1,1}{2,1}
\ncline{1,2}{2,2}
\ncline{2,1}{2,2}
\ncline{2,2}{2,3}
\ncline{2,3}{3,3}
\ncline{2,3}{3,2}   
\ncline{3,3}{3,2}
  \end{psmatrix}\]
\caption{Initial graph $G$}
\label{excep1}
\end{minipage}
\hfill
\begin{minipage}{0.34\linewidth}
\[\begin{psmatrix}[nodesep=3pt,rowsep=0.8cm,colsep=1cm] 
1&2&\\
3&4&5\\
&6&7
\ncline{1,1}{1,2}
\ncline{1,2}{2,1}
\ncline{1,1}{2,1}
\ncline{1,2}{2,2}
\ncline{2,1}{2,2}
\ncline[linecolor=green]{1,1}{2,2}
\ncline[linestyle=dashed, linecolor=red]{2,2}{2,3}
\ncline{2,3}{3,3}
\ncline{2,3}{3,2}   
\ncline{3,3}{3,2}
  \end{psmatrix}\]
\caption{Transformations on $G$}
\label{excep2}

\end{minipage}
\hfill
\begin{minipage}{0.32\linewidth}

\[\begin{psmatrix}[nodesep=5pt,rowsep=0.8cm,colsep=1cm] 
1&2&\\
3&4&5\\
&6&7
\ncline{1,1}{1,2}
\ncline{1,2}{2,1}
\ncline{1,1}{2,1}
\ncline{1,2}{2,2}
\ncline{2,1}{2,2}
\ncline{2,2}{1,1}
\ncline{2,3}{3,3}
\ncline{2,3}{3,2}   
\ncline{3,3}{3,2}
  \end{psmatrix}\]
\caption{Clustered graph $G'$}
\label{excep3}

\end{minipage}
\end{figure}

The solution presented in figures \ref{excep2} and \ref{excep3} can be written $(\{(4,5)\},\{(1,4)\})$ or $\{\{1,2,3,4\},\{5,6,7\}\}$ and has cost 2.
\end{example}

\vspace{0.35cm}
\paragraph{Weighted case} ~

\vspace{0.35cm}
\begin{definition}[weighted graph]
 
A weighted graph is an ordered pair $(V,w)$, where $V$ is the set of vertices of the graph and $w : \binom V 2 \to \bar{\mathbb R}=\mathbb R\cup \{-\infty,+\infty\}$ is the weight function. We consider that there is an edge between vertices $a$ and $b$ iff $w(a,b)\geqslant 0$. 
\end{definition}

Note that with this definition, even non-edges have a weight.

\vspace{0.35cm}
\begin{definition}[positive/negative part of a function]
 The positive and negative part of a function $f$, respectively written $f^+$ and $f^-$ are defined on the same domain as $f$ by $f^+(x)=\max (f(x),0)$ and $f^-(x)=\max(-f(x),0)$.
\end{definition}

The weighted optimal cluster editing problem still consists of transforming the initial graph $G=(V,w)$ into an unweighted clustered graph $G'$, but this time, the edges have a deletion/insertion cost given by their weight. Thus, the cost of inserting an edge in $G'$ between $a$ and $b$ is $w^-(a,b)$ and the cost of removing an edge is $w^+(a,b)$. Note that the unweighted case is a special case of the weighted problem where every edge has weight 1 and every non-edge has weight -1.

\subsubsection{State of the art}

The problem of cluster editing can be seen as a compromise between high edge-density community searching and low cut searching, two NP-hard problems, and as we can expect, the cluster editing problem itself has been proven NP-hard by Jan Kratochv\'{\i}l and Mirko Kriv{\'a}nek in \cite{NPC}. The problem is even APX-hard, that is, it is NP-hard to approximate the optimal solution with a given approximation factor $k>1$ \cite{APX}, and the best deterministic approximation factor reached today with a polynomial time algorithm is 2.5 \cite{2.5}. Deciding whether the cluster editing problem is NP-complete in the case of interval graphs is still an open question.

The cluster editing problem has numerous applications in various fields, like data processing, visualisation and bio-informatics. Returning to Clever, we were looking for a way to divide our reads into non-overlapping communities and cluster editing is a natural way to do so (note that clustered graphs are the only ones whose maximal cliques do not overlap). As for avoiding overlapping predictions, cluster editing is not completely sufficient: two clusters that do not overlap (\text{i.e.} that are vertex-disjoint) still may contain reads that overlap. However, this new model is more proper, reduces the number of overlapping predictions and therefore makes the post-processing step less damaging for the quality of the predictions. Moreover, this problem provides us a natural way to relax the clique definition (one missing edge, which can be due to a misread, can have decisive consequences on cliques, while clusters can support it). Cluster editing also has the advantage of being compatible with weighted graphs.
 
Indeed, each pair of reads will overlap more or less and be more or less close in size. It is therefore appreciable that the weighted cluster editing problem allows us to make some edges heavier than others, while clique enumerating requires a less adequate "all or nothing" model for edges.

However, cluster editing is a difficult problem to solve, even approximately, and the size of the biological data to be dealt with only allows linear time and space algorithms. We will therefore need large approximations to solve cluster editing while the clique enumerating problem still can be solved exactly on the read alignment graphs. 

There are already many heuristic cluster algorithms, well known examples are CAST \cite{cast}, CLICK \cite{click} or HCS \cite{hcs}, but the extreme size of the data and the special structure of the read alignment graphs, which allows solving clique enumerating in linear time, lead us to design a new specific algorithm.

\subsubsection{A special case: one dimensional point graph}

\vspace{0.35cm}
\begin{definition}[point graph]
\ 
 \begin{itemize}
  \itemb Given a set of points $V$ in an $n$-dimensional metric space $(M,d)$ and a threshold distance $l\in\mathbb R^+$, the induced point graph is the unweighted graph $G=\left(V,E=\{(a,b)\in\binom V 2 \mid d(a,b)\leqslant l\}\right)$
  \itemb An unweighted graph $G$ is said to be an $n$-dimensional point graph iff there exists a $n$-dimensional metric space $(M,d)$, a threshold distance $l$ and a set of points $V\subset M$ which induces $G$.
 \end{itemize}

\end{definition}

Given a one dimensional point graph, being able to place those vertices on a line provides us an order on the vertices (from left to right e.g.) and allows us to talk about consecutive vertices. The main result about cluster editing on point graphs is the following:

\vspace{0.35cm}
\begin{theorem}[Mannaa \cite{Mannaa}]
Given a one dimensional point graph, there exists an optimal clustering where all the clusters are made of consecutive vertices.
\label{Mannaa}
\end{theorem}

This result considerably reduces the size of the set of potential solutions to investigate and makes this set much easier to enumerate. Instead of enumerating the whole set of partitions of $V$, we only scan the points from left to right and have to determine whether we end the cluster or not after each vertex.

A dynamic programming algorithm \cite{Mannaa} follows: for $j\in\llbracket 0,|V|\rrbracket$, we call $\opt(j)$ the cost of optimally clustering vertices 0 to $j$ and for $j\in\llbracket 0,|V|\rrbracket$, $i\in\llbracket 0,j\rrbracket$, we call $\opt'(j,i)$ the cost of the best clustering for vertices 0 to $j$, where the last cluster has size $i+1$. We thus have:

\[\opt(j)=\left\{\begin{split}
&0\quad\text{ if }j=0\\
&\underset{0\leqslant i \leqslant j}{\min} \opt'(j,i)\text{ if }j>0
  \end{split}\right.\]

\[\opt'(j,i)=\left\{\begin{split}
&\opt(j-1)+\mathrm{deg}^+(v_j)\quad\text{ if }i=0\\
&\opt'(j-1,i-1)+|i-\mathrm{deg}^+(v_j)|\text{ if }0<i\leqslant j
  \end{split}\right.\]

Where $\mathrm{deg}^+(v)$ denotes the out degree of a vertex $v$ \textit{i.e.} the number of arcs starting at $v$.

We can compute all $\opt'$ value in $O(n^2)$ time and space:

\vspace{0.35cm}
\begin{algorithm}[H]

$\opt'(0,0)\leftarrow 0$

\For {$j$ from 1 to $n-1$}{

$\opt'(j,0)\leftarrow\deg^+(v_j)+\underset{0\leqslant i\leqslant j-1}{\min} \opt'(j-1,i)$

\For {$i$ from 1 to $j$}{

$\opt'(j,i)\leftarrow|i-\deg^+(v_j)|+\opt'(j-1,i-1)$
}
}
\caption{Computing the values of $\opt'$}
\end{algorithm}

We then know which cluster to make by searching where the function $\opt'(j,\cdot)$ reaches its minimum. Once we know $\opt'$, it can be processed in $O(n)$.

\paragraph{Application} ~

As previously said, our read alignment graphs fail to be interval graphs because of size differences between the reads: if the size is too different, there will not be an edge between the two vertices no matter how much they overlap. If we only take into account reads whose sizes are close enough to allow edges, we would have an interval graph, but, as NP-completeness of cluster editing on interval graphs still is an open problem, there is no guarantee that interval graphs would be of any help. But by considering only reads whose sizes are close, we end up having not only an interval graph, but a one dimensional point graph whose points set is for example the left endpoint of the intervals, and whose threshold distance is their length. If we sort the reads by their left endpoint, we would then be able to use the 1D point graph algorithm of Mannaa.

\section{Contribution}

\subsection{Algorithm on weighted point graphs}

\subsubsection{Exact algorithm}

\vspace{0.35cm}
We define a weighted point graph as follows:

\vspace{0.35cm}
\begin{definition}[weighted point graph]
\ 
 \begin{itemize}
  \itemb Given a set of points $V$ in an $n$-dimensional metric space $(M,d)$ and a decreasing function $f:\mathbb R^+\to\bar{\mathbb R}=\mathbb R\cup \{-\infty,+\infty\}$, the induced point graph is the weighted graph $G=(V,w:(a,b)\mapsto f(d(a,b)))$
  \itemb A weighted graph $G$ is said to be an $n$-dimensional weighted point graph iff there exists a $n$-dimensional metric space $(M,d)$, a decreasing function \mbox{$f:\mathbb R^+\to\bar{\mathbb R}$} and a set of point $V\subset M$ which induce $G$.
 \end{itemize}

\end{definition}

\begin{remark} 
\ 
\begin{itemize}
 \itemb An unweighted point graph is a special case of weighted point graph with \[\fonction f {\mathbb R^+} {\bar{\mathbb R}} a 
{\left\{\begin{split}
         1\quad&\text{if }a\leqslant l\\
-1\quad&\text{if } a>l
        \end{split}\right.}\]
\itemb Suppose now we create a weighted read alignment graph with only reads of the same size. The closer the left endpoint of two reads will be, the more they will overlap and the heavier the edge will be. If all the reads have the same size, weighted read alignment graphs are one dimensional weighted point graphs, like in the unweighted case.
\end{itemize}

\end{remark}

With this definition, the proof of theorem \ref{Mannaa} (see \cite{Mannaa}) still holds:

\vspace{0.35cm}
\begin{theorem}
Given a one dimensional weighted point graph, there exists an optimal clustering where all the clusters are made of consecutive vertices.
\label{wmannaa}
\end{theorem}

We now have:

\[\opt(j)=\left\{\begin{split}
&0\quad\text{ if }j=0\\
&\underset{0\leqslant i \leqslant j}{\min} opt'(j,i)\text{ if }j>0
  \end{split}\right.\]

\[\opt'(j,i)=\left\{\begin{split}
&\opt(j-1)+\dsum{k=0}{j-1} w^+(j,k)\quad\text{ if }i=0\\
&\opt'(j-1,i-1)+\dsum{k=0}{j-1} w^+(j,k)+\dsum{k=j-i}{j-1} w^-(j,k)\text{ if }0<i\leqslant j
  \end{split}\right.\]

The formula for each of the $O(n^2)$ values of $\opt'$ now includes a sum that takes $O(n)$ time to compute. Still, those sums are often alike and by computing only one of them, we can deduce the next one in constant time and we still end up with an $O(n^2)$ algorithm:

\vspace{0.35cm}
\begin{algorithm}[H]

$\opt'(0,0)\leftarrow 0$ 

\For {$j$ from 1 to $n-1$}{

$X\leftarrow \underset{0\leqslant k\leqslant j-1}{\sum} w^+(j,k)$

$\opt'(j,0)\leftarrow X+\underset{0\leqslant i\leqslant j-1}{\min} \opt'(j-1,i)$

\For {$i$ from 1 to $j$}{

$X\leftarrow X-w(j,j-i)$

$\opt'(j,i)\leftarrow X+\opt'(j-1,i-1)$
}
}
\caption{Computing the values of $\opt'$ in a weighted graph}
\label{walgo}
\end{algorithm}

\begin{remark}
 If we do not initialize $X$ with $\underset{0\leqslant k\leqslant j-1}{\sum} w^+(j,k)$, $\opt(j)$ will no longer give the cost of the optimal clustering for vertices 0 to $j$. However, $X$ is nothing but an additive constant in $\opt'(j,\cdot)$ and what we are really interested in is to determine where the minimum of $\opt'(j,\cdot)$ is reached. If we initialize $X$ with any other value, for example, zero, we will still find the optimal clustering.
\label{X}
\end{remark}

\subsubsection{Heuristic algorithms}

When the number of nodes of the input graph reaches hundreds of millions, even $O(n^2)$ algorithms are not suitable anymore. However, we can be much faster with an acceptable approximation by noticing the following: if the optimal clustering for nodes 0 to $j$ is reached with a last cluster of size $i$, chances are very low that the optimal clustering for nodes 0 to $j+1$ is reached with a last cluster of size strictly bigger than $i+1$ and we do not need to compute $\opt'(j+1,k)$ for $k\in\llbracket i+2,j+1\rrbracket$. There are of course counter examples, especially with small clusters, but this approximation is generally safe when clusters exceed hundreds of vertices. 

We therefore obtain a first heuristic algorithm by making the following change on line 5 of the exact algorithm:

\vspace{0.35cm}
\For {$i$ from 1 to $\argmin(\opt'(j,\cdot))+1$}

\vspace{0.35cm}

We can however, make our heuristic more accurate by noticing that most of the time when errors happen, there was still a positive edge between the vertex $j$ and the vertex $j-i$. 

Line 5 of the second algorithm is:

\vspace{0.35cm}
\For {$i$ from 1 to $\max\left(\argmin(\opt'(j,\cdot))+1, \max \left\{i\mid w(j,j-i)>0\right\}\right)$}

\vspace{0.35cm}
If we initialize $X$ with 0 as suggested in note \ref{X}, our two algorithms now run in $O(n\*k)$ where $k$ is the size of the clusters (we will not, however, be able to determine the cost of the provided solution in less than $O(n^2)$). Finally, we can note that to determine the clustering, all we need to know at the end are the $\argmin(\opt'(j,\cdot))_{0\leqslant j\leqslant n}$. Once we are done computing $\opt'(j,\cdot)$, we then can erase the values of $\opt'(j-1,\cdot)$ as long as we keep memory of $\argmin(\opt'(j-1,\cdot))$. This reduces the space complexity of the algorithm from $O(n^2)$ to $O(n)$.

\subsubsection{Experimental results}

Our tests generate random weighted graphs of size $n$ by placing $n$ points randomly on $[0,1]$. We then choose a threshold distance $l$, the distance between two points $a$ and $b$ is of course $|a-b|$ and we create the graph by composing the distance with the function:
\[\fonction {f_l} {\mathbb R^+} {\bar{\mathbb R}} a {\frac{l^2-a^2}{la}}\]

This function was chosen because of three intuitive properties we expected from the weight: $f_l(0)=+\infty$, $f_l(l)=0$ and $\underset{x\to+\infty}{\lim}f_l(x)=-\infty$.

The values of the threshold distance $l$ will determine the edge density of the graph and therefore the size $c$ of the clusters. All the results in the following table are average results on 1,000 measures on graphs of 10,000 vertices. Those results aim at showing the speed (through the number of $\opt'$ values it computes) and the accuracy (through the cost of the answer it returns) of both heuristics but also the impact of the size of the clusters.

\vspace{0.35cm}
\underline{$l=0.1$:}

average number of clusters (exact solution): 6.121

average size of clusters: 1633.7

\begin{center}
\begin{tabular}{|c|c|c|}
\hline
& Cost of the solution & \begin{minipage}{0.22\linewidth}Number of $\opt'$ values calculated\vspace{0.05cm}\end{minipage}\\
\hline
Exact algorithm & 5,269,293.579 & 49,995,000.000\\
\hline
Second heuristic algorithm & 5,269,293.579  & 14,734,267.491 \\
\hline
First heuristic algorithm & 5,269,293.579 &  14,733,439.917\\
\hline
\end{tabular}
\end{center}

\vspace{0.35cm}

\underline{$l=0.01$:}

average number of clusters (exact solution): 62.809

average size of clusters: 159.21

\begin{center}
\begin{tabular}{|c|c|c|}
\hline
& Cost of the solution & \begin{minipage}{0.22\linewidth}Number of $\opt'$ values calculated\vspace{0.05cm}\end{minipage}\\
\hline
Exact algorithm & 535,093.568 & 49,995,000.000\\
\hline
Second heuristic algorithm & 535,094.783  & 1,525,308.148 \\
\hline
First heuristic algorithm & 535,283.994 &  1,515,427.532\\
\hline
\end{tabular}
\end{center}

\vspace{0.35cm}
\underline{$l=0.001$:}

average number of clusters (exact solution): 644.327

average size of clusters: 15.520

\begin{center}
\begin{tabular}{|c|c|c|}
\hline
& Cost of the solution & \begin{minipage}{0.22\linewidth}Number of $\opt'$ values calculated\vspace{0.05cm}\end{minipage}\\
\hline
Exact algorithm & 40,964.600 & 49,995,000.000\\
\hline
Second heuristic algorithm & 40,968.290  & 139,936.991 \\
\hline
First heuristic algorithm & 41,378.066 &  132,794.486\\
\hline
\end{tabular}
\end{center}

\vspace{0.35cm}
We can see that the number of calculations the heuristics do is proportional to the size of the clusters. As it was mentioned previously, the chances that errors are made are getting very low when the clusters grow bigger. We can also see that for the price of few additional calculations, the second heuristic avoids most of the mistakes made by the first one.

\subsection{Back to the read alignment graph}

\subsubsection{Comparison with the one dimensional case}

As knowing the left endpoint and the length of every read gives us the required information to determine the weights of the edges, we can place our reads in a two-dimensional plane and look for a metric and a weight function that would make it a point graph. However,  dynamic programming was only possible because of the very special structure of the set of possible solutions theorem \ref{wmannaa} leaves to investigate. As this theorem no longer holds in two dimensions, there is no easy way to adapt the one dimensional point graph algorithm.

Still, we saw that in most of the cases, the weight of the edges will be determined by the position of the reads and not by the size difference. Then, even though we are in two dimensions, points are much more staggered over one dimension than over the other and there is still hope that our one dimensional study helps.

\subsubsection{Best clustering for a given order}

Looking back at algorithm \ref{walgo}, we realise that it solves a more general problem than just optimal clustering for one dimensional point graphs: given any kind of graph and an order on the vertices, it will find the optimal clustering where clusters only contain consecutive vertices. Theorem \ref{wmannaa} only ensures that an optimal clustering with the left-to-right order is a general optimal clustering.

Regarding the first heuristic algorithm, if two vertices with a very heavy edge are far from each other in the order, the algorithm may not even try to put them in the same clusters and will pay a high cost for pulling those vertices apart. However, this will only happen if the order is bad, but with good orders, the first heuristic algorithm will return a good solution too and will still be decisively faster than the exact one.

The small refinement brought by heuristic two makes no sense in our case since a positive edge between the vertex $j-i$ and the vertex $j$ does not necessarily imply a positive edge between the vertex $j-i$ and the vertex $j-1$ and the heuristic may lead us to calculate $\opt'(j,i)$ without even knowing $\opt'(j-1,i-1)$.

In what follows, we will therefore use the first heuristic to address this problem.

\subsubsection{Best order for clustering}

This leads us to studying a new problem: the optimal order for cluster editing. 

Given an order on the vertices, we define its cost as the cost of the optimal clustering where clusters are made of consecutive vertices. The purpose is now to find a fast heuristic for finding the optimal order and to combine it with the previous heuristic for best clustering for a given order.

\vspace{0.35cm}
First of all, let us see why the previous approach no longer works. Imagine we have 10 reads stemming from two different alleles (reads 1 to 5 from the first one and read 6 to 10 from the other) and that the position of the left endpoint of the reads on the reference genome happen to be close even if the reads stem from different alleles. If reads stemming from different alleles have different sizes, we should still be able to distinguish them, and this is precisely what we expect from our cluster editing algorithm.

\begin{figure}[!h]
\centering\[\begin{pspicture}(6.5,4)
\rput(6.1,0.2){\small{\text{position of}}}
\rput(6.1,-0.2){\small{\text{left endpoint}}}
\psline{-}(0,0)(5.2,0)
\psline[linestyle=dashed](5.2,0)(6.4,0)
\psline{->}(6.4,0)(6.5,0)
\rput(-0.3,4.2){\small{\text{size}}}
\psline{->}(0,0)(0,4.3)

\rput(1,0.9){\small 1}
\psdot(1.1,0.7)
\psline[linestyle=dashed, linecolor=gray](1.1,0)(1.1,0.7)
\rput(1.9,1.2){\small 2}
\psdot(2,1)
\psline[linestyle=dashed, linecolor=gray](2,0)(2,1)
\rput(3.4,0.7){\small 3}
\psdot(3.5,0.5)
\psline[linestyle=dashed, linecolor=gray](3.5,0)(3.5,0.5)
\rput(4.1,1){\small 4}
\psdot(4.2,0.8)
\psline[linestyle=dashed, linecolor=gray](4.2,0)(4.2,0.8)
\rput(5,1.3){\small 5}
\psdot(5.1,1.1)
\psline[linestyle=dashed, linecolor=gray](5.1,0)(5.1,1.1)

\rput(0.5,3.7){\small 6}
\psdot(0.6,3.5)
\psline[linestyle=dashed, linecolor=gray](0.6,0)(0.6,3.5)
\rput(1.4,4.2){\small 7}
\psdot(1.5,4)
\psline[linestyle=dashed, linecolor=gray](1.5,0)(1.5,4)
\rput(2.8,3.9){\small 8}
\psdot(2.7,3.7)
\psline[linestyle=dashed, linecolor=gray](2.7,0)(2.7,3.7)
\rput(3.6,3.8){\small 9}
\psdot(3.8,3.6)
\psline[linestyle=dashed, linecolor=gray](3.8,0)(3.8,3.6)
\rput(4.6,3.9){\small 10}
\psdot(4.8,3.7)
\psline[linestyle=dashed, linecolor=gray](4.8,0)(4.8,3.7)
\end{pspicture}\]
\caption{Example of flaw in the left-to-right approach}
\label{left-right}
\end{figure}
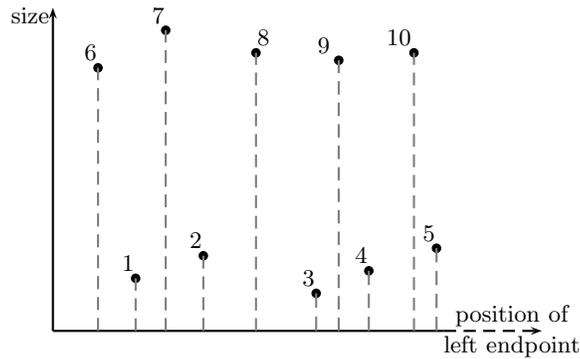

Figure \ref{left-right} illustrates this example; reads are organised on a plane according to their positions and lengths. Such a configuration of the reads is very likely to happen in practice.

Here, we should look for optimal clustering with vertices 1, 2, 3, 4 and 5 and with vertices 6, 7, 8, 9 and 10 separately . However, the left-to-right order is [6,1,7,2,8,3,9,4,10,5] and the arising clustering would clearly be sub-optimal.

\vspace{0.35cm}
A good way to solve this problem is the following: start with the left vertex, and at each step, insert in the order the unassigned vertex which is the closest from the previous one (\textit{i.e.} the one with the heaviest edge). In the previous example, the resulting order would be $[6,7,8,9,10,5,4,3,2,1]$. There will be $n$ steps and at each step, the program has to investigate the $O(k)$ vertices whose associated reads overlap with the read of the previous vertex (the edges we do not investigate have a $-\infty$ weight). If all such vertices are already assigned, choose arbitrarily which vertex will follow in the order.

This algorithm is quite simple and gives significantly better results than the previous one, but it still is far from being flawless.

The reads configuration depicted in figure \ref{closest} highlights a typical problem:

\begin{figure}[!h]
\centering\[\begin{pspicture}(6,4)
\rput(6.1,0.2){\small{\text{position of}}}
\rput(6.1,-0.2){\small{\text{left endpoint}}}
\psline{-}(0,0)(5.2,0)
\psline[linestyle=dashed](5.2,0)(6.4,0)
\psline{->}(6.4,0)(6.5,0)
\rput(-0.3,4.2){\small{\text{size}}}
\psline{->}(0,0)(0,4.3)

\rput(0.6,1.9){\small 1}
\psdot(0.8,1.8)
\rput(1.4,1.6){\small 2}
\psdot(1.6,1.5)
\rput(2.4,1.9){\small 3}
\psdot(2.6,1.8)
\rput(2.8,2.1){\small 4}
\psdot(3,2)

\rput(3,1.3){\small 5}
\psdot(3.2,1.2)
\rput(3,2.5){\small 6}
\psdot(3.2,2.4)
\rput(3.6,2.1){\small 7}
\psdot(3.8,2)

\rput(4,2.3){\small 8}
\psdot(4.2,2.2)
\rput(5,1.7){\small 9}
\psdot(5.2,1.6)

\end{pspicture}\]
\caption{Example of flaw in the closest-neighbour approach}
\label{closest}
\end{figure}
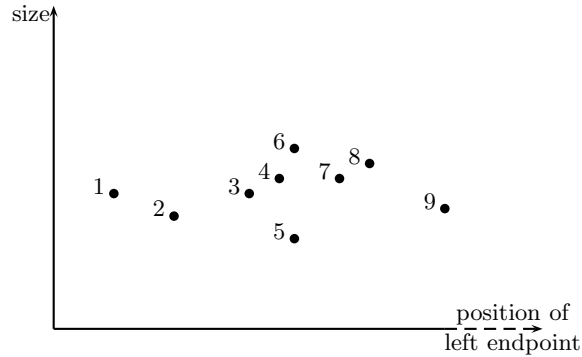

If we apply the previous algorithm with the example of figure \ref{closest}, the resulting order will be $[1,2,3,4,6,7,8,9,\cdots]$ and the vertex 5 will be skipped. The algorithm will go back to vertex 5 only long after and vertex 5 will probably end up in a singleton.

A generalisation of this process is to choose $h\geqslant 1$ (3 or 5 already give good results) and to look at the unassigned vertex that maximises the sum of the weight of the edges with the $h$ previously assigned vertices. The complexity is now $O(hkn)=O(kn)$, just like the clustering algorithm we will run on the provided order.

Now, let us see what will happen when, in the clustering, we find for the $j$ first vertices, the vertex $j$ ends up in a singleton. In this case, we know before choosing which vertex will be at rank $j+1$ that the vertex $j$ and $j-1$ will not end up in the same cluster in the final clustering (our heuristic will not even investigate those solutions). Thus, it makes no sense to take into account the weight of the edges between the vertex $j-1$ and the candidates for rank $j+1$ as none of those vertices  will end up in the same cluster than $j-1$. 

In our final algorithm, we will run simultaneously the heuristic for best order and for best clustering with a given order: as soon as we know the vertex $j$, we compute $opt'(j,\cdot)$ and we will use $h=\argmin(opt'(j,\cdot))$ to choose the vertex $j+1$. Our algorithm runs in $O(n\*k^2)$ which is acceptable since $k$ can be assumed constant.

\vspace{0.35cm}

\subsubsection{Final cluster editing algorithm}

We are given a set of vertices that we number from 0 to $n-1$ from left to right, and a weight function. Computing once every value of $w$ for non overlapping reads can be done in $O(nk)$. We then can store those values in a hashtable to determine $w(a,b)$ in constant time. If the pair $(a,b)$ has no entry in the hashtable, we know that the reads do not overlap and that $w(a,b)=- \infty$. 

We then create the following objects:
\begin{compactitem}
 \itemb an integer array \texttt{order} of size $n$. Ultimately, $\order(i)$ will be the $i^{\text{th}}$ vertex of the order we will use for cluster editing. 
 \itemb a boolean array \texttt{assigned} of size $n$ initialized with false. It will indicate whether vertex $i$ has already been assigned in $\order$.
 \itemb an (integer list) array \texttt{neighbours} of size $n$, such that $\neighbours(a)$ contains a vertex $b$ iff their associated reads overlap. In other words, $\neighbours(a)$ is the list of vertices $b$ such that $w(a,b) \neq - \infty$. We can also create this array and the weight function $w$ simultaneously.
 \itemb functions \texttt{argmin} and \texttt{min}: given an integer list $l$ and a function $f:\mathbb N\to\mathbb R$, they respectively return the element of $l$ that minimises $f$ and the values of the minimum.
 \itemb functions \texttt{argmax} and \texttt{max}: given an integer list $l$ and a function $f:\mathbb N\to\mathbb R$, they respectively return the element of $l$ that maximises $f$ and the value of the maximum.
 \itemb a function \texttt{select}: given an integer list $l$ and a predicate $p:\mathbb N\to\{\mathrm{true},\mathrm{false}\}$, it returns the list of all elements of $l$ that satisfy the predicate $p$.
 \itemb A function \texttt{anyvertex} which returns a vertex that has not been assigned yet. A good thing to do is to return the unassigned vertex with the lowest index and to keep memory of the last vertex the function returned so that you know where to start the investigation next time the function is called. It is correct as you will never unassign a previously assigned vertex and guarantee that no matter how many times you use $\anyvertex$, the total computation time will always be $O(n)$.
 \itemb A matrix \texttt{opt'} of size $n\*n$ and an array \texttt{opt} of size $n$ whose purpose will be the same as previously. We will also use an array \texttt{size} of size $n$ to keep $\argmin(opt'(j,\cdot))$ in memory. Thus, $\size(j)$+1 will indicate the size of the last cluster in the optimal clustering of the first $j$ vertices. 
\end{compactitem}

\begin{algorithm}[H]

$\opt'(0,0)\leftarrow 0$

$\opt(0)\leftarrow 0$

$\size(0)\leftarrow 0$

$\order(0)\leftarrow 0$

$\assigned(0)\leftarrow \true$

\For {$j$ from 1 to $n-1$}{

$X\leftarrow 0$

$\opt'(j,0)\leftarrow X+\underset{0\leqslant i\leqslant j-1}{\min} \opt'(j-1,i)$

$\mathrm{candidates}=\select(\neighbours(order(j-1)),p:i\mapsto not(\assigned(i)))$

\eIf{$\mathrm{candidates}=[\quad \!\!\!\!]$}
{$\order(j)\leftarrow \anyvertex()$}
{$\order(j)\leftarrow \argmax(\mathrm{candidates},f:i\mapsto\dsum{k=0}{\size(j-1)}w(i,\order(j-1-k))$}

$\assigned(\order(j))\leftarrow \true$

\For {$i$ from 1 to $\size(j-1)$}{

$X \leftarrow X-w(j,j-i)$

$\opt'(j,i)\leftarrow X+\opt'(j-1,i-1)$
}

$\opt(j)\leftarrow \min(\llbracket 0,\size(j-1)\rrbracket,f:i\mapsto \opt'(j,i))$

$\size(j)\leftarrow \argmin(\llbracket 0,\size(j-1)\rrbracket,f:i\mapsto \opt'(j,i))$
}

$\clusters\leftarrow [\quad \!\!\!\!]$

$j\leftarrow n-1$

\While{$j>0$}{

$\clusters\leftarrow\clusters \cup \left\{\dcup{k=j-\size(j)} j \order(k)\right\}$

$j\leftarrow j-\size(j)-1$

}

\Return{$\clusters$}
\caption{Final cluster editing algorithm}
\end{algorithm}

As we said previously, it is possible to erase $\opt'(j-1,\cdot)$ after the for loop from line 14 to 16. To avoid creating an $n\*n$ array, one can also create a $2\*n$ array and replace all the $\opt'(i,j)$ in the algorithm with $\opt'(i\!\!\! \mod 2, j)$.

The final algorithm runs in $O(nk)$ in time and $O(n)$ in space.

\subsection{Experimental Results}

\subsubsection{Protocol}

To compare our new algorithm with existing state-of-the-art tools, we ran experiments based on Craig Venter's genome \cite{Venter} and used UCSC's Simseq to simulate reads of mean size $\mu=112$ and standard deviation $\sigma=15$, in accordance to biological datasets. The reference genome was downloaded from UCSC genome browser. The main advantage of using simulated data is that we know exactly what are the insertions and deletions we expect our program to find while real data would highlight discordance between existing tools but would not be able to determine which ones give the best results.

The test consists of searching insertions and deletions on the first chromosome of the genome. It is the largest one and the number of insertions and deletions is large enough to draw relevant conclusion about the quality of the predictions. We compare the results provided by our new tool to the one provided by the former clique enumerating version of Clever and to the other existing state-of-the-art tools: GASV \cite{GASV}, Variation Hunter \cite{VH}, Breakdancer \cite{Breakdancer}, Hydra \cite{Hydra}, MoDIL \cite{MoDIL}, Pindel \cite{Pindel} and SV-seq2 \cite{SV-seq2}. Note that GASV and SV-Seq2 only predict deletions.

We compared those tools on two criteria: the precision of the predictions, \textit{i.e.} the rate of predictions that were actual insertions and deletions, and recall, \textit{i.e.} the rate of insertions and deletions that were predicted. We also indicate the number of exclusive predictions each tool makes \textit{i.e.} the number of actual variations only this tool detects. Note that the proximity between the two versions of Clever will tend to lower their rates of exclusive predictions.
We computed scores separately on insertions and deletions and we distinguished 5 length ranges.

\subsubsection{Results}

The graph that arises from this dataset contains $n=41,437,854$ vertices and the number of edges whose weight is not $-\infty$ is $nk=588,570,096$. Our program was given the files containing the vertices and the edges, which are respectively 3.3GB and 15GB. An implementation in ocaml of the main algorithm (including the creation of the weight hashtable) ran in 55 minutes on a single core and peaked at 72.2GB. The interpretation of the results leading from the clustering to the predictions took less than 20 minutes. With the other tools, the computation could last up to several hundreds of hours.

Results we present in this report were obtained with a threshold of 0.4 for the weight of the edges and a false discovery rate of 0.1 when applying Benjamini-Hochberg process to the $p$-values file. We also used a post-process file to eliminate overlapping predictions.

Comparative results of all the tools we tested are presented in the
following table.

\begin{center}
\begin{longtable}{|c||c|c|c||c|c|c|}
\hline
&\multicolumn{3}{|c||}{Insertions}&\multicolumn{3}{|c|}{Deletions}\\
\hline
\hline
\endhead
\multicolumn{7}{|c|}{Length range: 20-49 - 644 true insertions - 618 true deletions}\\
\hline
Tool&Precision&Recall&Excl.&Precision&Recall&Excl.\\
\hline
Clever (cluster)&86.7&50.0&0.5&65.2&63.8&0.2\\
\hline
Clever (cliques)&93.4&46.7&0.6&95.0&65.0&0\\
\hline
BreakDancer&--&7.0&0&89.7&7.4&0\\
\hline
GASV&-&-&-&9.3&44.7&0.2\\
\hline
HYDRA&0&0&0&--&0.2&0\\
\hline
VH&52.8&10.4&0&83.1&13.3&0\\
\hline
PINDEL&95.3&67.9&18.3&68.8&76.1&4.9\\
\hline
SV-seq2&-&-&-&--&0&0\\
\hline
MoDIL&62.4&61.8&5.3&73.4&80.3&2.9\\
\hline
\hline
&\multicolumn{3}{|c||}{Insertions}&\multicolumn{3}{|c|}{Deletions}\\
\hline
\multicolumn{7}{|c|}{Length range: 50-99 - 153 true insertions - 125 true deletions}\\
\hline
Tool&Precision&Recall&Excl.&Precision&Recall&Excl.\\
\hline
Clever (cluster)&53.8&79.1&0&55.2&77.6&0.8\\
\hline
Clever (cliques)&67.0&73.2&0&69.2&76.8&0\\
\hline
BreakDancer&93.7&57.5&0&96.8&51.2&0\\
\hline
GASV&-&-&-&57&43.2&0.8\\
\hline
HYDRA&0&0&0&--&4.8&0\\
\hline
VH&60.4&80.4&0&76.4&73.6&0\\
\hline
PINDEL&88.9&27.5&0&74.4&37.6&0\\
\hline
SV-seq2&-&-&-&--&0&0\\
\hline
MoDIL&50.9&92.2&5.9&55.3&84.8&5.6\\
\hline
\hline
\multicolumn{7}{|c|}{Length range: 100-249 - 90 true insertions - 63 true deletions}\\
\hline
Tool&Precision&Recall&Excl.&Precision&Recall&Excl.\\
\hline
Clever (cluster)&80.0&27.8&0&68.8&57.1&1.6\\
\hline
Clever (cliques)&50.0&27.8&1.1&77.5&54.0&0\\
\hline
BreakDancer&49&25.6&3.3&63.7&49.2&0\\
\hline
GASV&-&-&-&100&42.9&1.6\\
\hline
HYDRA&0&0&0&78.0&42.9&0\\
\hline
VH&50&52.2&3.3&36.7&60.3&3.2\\
\hline
PINDEL&--&3.3&0&65.0&15.9&0\\
\hline
SV-seq2&-&-&-&--&0&0\\
\hline
MoDIL&40.6&55.6&7.8&58.8&34.9&3.2\\
\hline
\hline
\multicolumn{7}{|c|}{Length range: 250-999 - 106 true insertions - 113 true deletions}\\
\hline
Tool&Precision&Recall&Excl.&Precision&Recall&Excl.\\
\hline
Clever (cluster)&--&0&0&92.9&69.9&0\\
\hline
Clever (cliques)&--&0&0&91.9&70.8&0\\
\hline
BreakDancer&--&0&0&71.8&68.1&0\\
\hline
GASV&-&-&-&0.5&61.1&0\\
\hline
HYDRA&--&0&0&82.1&73.5&1.8\\
\hline
VH&--&0&0&95.2&71.7&0.9\\
\hline
PINDEL&--&0&0&94.8&62.8&0\\
\hline
SV-seq2&-&-&-&--&0&0\\
\hline
MoDIL&--&0&0&--&0&0\\
\hline
\hline
\multicolumn{7}{|c|}{Length range: 1000-50000 - 27 true insertions - 21 true deletions}\\
\hline
Tool&Precision&Recall&Excl.&Precision&Recall&Excl.\\
\hline
Clever (cluster)&--&0&0&77.8&66.7&0\\
\hline
Clever (cliques)&--&0&0&87.5&66.7&0\\
\hline
BreakDancer&--&0&0&76.5&61.9&0\\
\hline
GASV&-&-&-&66.7&66.7&0\\
\hline
HYDRA&--&0&0&37.2&66.7&0\\
\hline
VH&--&0&0&70.0&66.7&0\\
\hline
PINDEL&--&0&0&60.0&57.1&0\\
\hline
SV-seq2&-&-&-&--&0&0\\
\hline
MoDIL&--&0&0&--&0&0\\
\hline
\end{longtable}
\end{center}

The first thing we can notice is that although the new version of Clever uses a heuristic algorithm, the predictions are comparable to the ones the previous version provided, with even some exclusive predictions. This model also gives good results for every length range, for both insertions and deletions. Moreover, our predictions are as good as the predictions of considerably slower tools, giving hope that Clever can be used to investigate data that the other tools could not handle.

\section{Conclusion}

In this report, we introduced a new model based on cluster editing for next-generation sequencing. This model has many advantages, helps relax the definition of cliques and is an essential step toward algorithms which avoid overlapping predictions. However, the underlying problem is very difficult and requires large approximation to be solved on real biological data while the previous approach was based on enumerating cliques, a problem that could be solved exactly. We therefore introduced a $O(nk^2)$ time and $O(nk)$ space heuristic for cluster editing where $k$ is a bound on the size of the clusters and can be assumed constant. The resulting tool provides good predictions for both insertions and deletions of various length and can stand up to comparison with other state-of-the-art tools. Moreover, there is surely considerable room for improvements, in the heuristic for example, and there is hope that the results of this method improve faster and ultimately get decisively better than the results provided by 
the clique enumerating method which is already the results of years of work and use exact algorithms. 

In addition, besides the fact that it helped validate the clustering model, the $O(nk^2)$ heuristic for cluster editing can be seen as a contribution in itself, as the problems it solves have many applications in various fields.

Finally, the fact that our algorithm runs in $O(nk^2)$ makes it possible to deal with high coverage data, which will be more and more common in the coming years.

%
\subsection*{Acknowledgements}
The authors thank COST action BM1006 SeqAhead for
supporting this work and Murray Patterson for his advice on an early
version of this manuscript.


\bibliographystyle{plain}
\bibliography{biblio}


\end{document}